\documentstyle[preprint,aps]{revtex}

\preprint{IMSc-99/08/30A}

\begin{document}
\draft 
\title{Solar Neutrinos and the Eclipse Effect}
\author
{Mohan Narayan,${}^a$ G. Rajasekaran,${}^a$ Rahul Sinha ${}^a$ 
and C.P. Burgess ${}^b$} 
\address
{${}^a$ Institute of Mathematical Sciences, Madras 600 113, India.~\\
${}^b$ Physics Department, McGill University, 
Montr\'eal, Qu\'ebec, Canada, H3A 2T8.}
\maketitle

\begin{abstract}

The solar neutrino counting rate in a real time detector like
Super--Kamiokanda, SNO, or Borexino is enhanced due to neutrino
oscillations in the Moon during a partial or total solar eclipse. The
enhancement is calculated as a function of the neutrino parameters in
the case of three flavor mixing. This enhancement, if seen, can further
help to determine the neutrino parameters.
 
\end{abstract}

\newpage
\section{Introduction}

The Sun is a copious source of neutrinos with a wide spectrum of
energies and these neutrinos have been detected by terrestrial
neutrino detectors, although at a rate lower than expected from
theoretical calculations.  A new generation of detectors
Super-Kamiokanda, SNO and Borexino \cite{totsuka,sno,raghavan} with high counting rates will soon be
producing abundant data on solar neutrinos; among these
Super-Kamiokanda has already started producing results \cite{suzuki}.  Mixing and the consequent
oscillations among the neutrinos of different flavors is generally
believed to be the cause of the reduced intensity of neutrino flux
detected on Earth. However, neutrino--oscillation is a complex
phenomenon depending on many unknown parameters (six parameters for
three flavors $\nu_{e}$, $\nu_{\mu}$, $\nu_{\tau}$) and considerable
amount of experimental work and ingenuity will be required before the
neutrino problem is solved.

Hence it would be desirable, if  apart from direct detection, we can
subject the 
solar neutrino beam to further tests by passing it through different
amounts of matter, in our attempts to learn more about the neutrinos. Nature
has fortunately provided us with such opportunities:
$(1)$ Neutrinos detected at night pass through Earth.
$(2)$ Neutrinos detected during a solar eclipse pass through the Moon.
$(3)$ Neutrinos detected at the far side of Earth during a solar
eclipse pass through the Moon and Earth. We shall call this
scenario $(3)$ a
{\em double eclipse}. The scenario $(1)$ has been studied in the literature
rather extensively \cite{blwz,ours}. The purpose of the present work is to examine
the scenarios $(2)$, and $(3)$. Previous works \cite{currsc,ourecl1,Sher} have
discussed scenario $(2)$, however they are incomplete in many
respects.

The plan of the paper is as follows. The relevant astronomy
is presented in Sec.\ref{ecl}. In Sec.\ref{th} we give the theory of the passage of the
solar neutrinos through the Moon and Earth, taking into account properly
the non-adiabatic transitions occuring at the boundaries of the Moon
and Earth. In Sec.\ref{calc} we present the
numerical calculations of the neutrino detection rates during the single and
double eclipses and the results. Sec.\ref{disc} is devoted to discussion.

\section{Eclipses and double eclipses}
\label{ecl}

Solar neutrinos are produced within the solar core whose radius is of
order $1/10$ of the solar radius and we shall approximate this by a point
at the center of the Sun. What is required for our purpose is that the lunar
disc must cover this point at the center of the Sun and so as far as the
neutrino radiation is concerned , the solar eclipse is more like an 
occultation of a star or a planet by the Moon.

Astronomers characterize the solar eclipse by the optical coverage $C$
which is defined as the ratio of the area of the solar disc covered by 
the lunar disc to the total area of the solar disc. We first calculate
how $C$ varies with time during an eclipse.

There are three parameters which completely characterize the
time-dependence of any solar eclipse. 
\begin{itemize}
\item
The first parameter is the ratio, $a$, of
the apparent radii (in radians, say) of the Moon and Sun's discs as seen from 
Earth, $a = r_m/r_s$.  The eclipse is total if $a \ge 1$ and it is
annular if $a < 1$. 
\item
The second parameter is the duration of the eclipse, $T$, 
measured as the length of time between the initial and final instants where the
Moon's and Sun's discs just touch.
\item
Finally, we have the impact parameter, $p = \ell/r_s$, defined as
the apparent distance of closest approach between the center of the Sun's
disc and the center of the Moon's disc, in units of the Sun's apparent radius.
\end{itemize}

Given these quantities, the expression for $C$ as a function of time
$t$ during an eclipse is determined by simple geometry as
follows. First it is convenient to change the independent variable
from $t$ to $d$, defined as the apparent separation between the
centers of the Sun and the Moon (in units of the apparent solar
radius).  If $t=0$ represents the point of maximum coverage, which
occurs when $d$ equals $p$ and is minimum and $t = \mp T/2$ correspond
to the start and end of the eclipse when $d = 1 + a$, then $d$ is
given by
\begin{equation}
\label{eqdoft}
d(t) = \left\{p \left[1- \left({2 t \over T}\right)^2 \right] +
\left( 1 + {a} \right)^2 \left( {2 t \over T} \right)^2
\right\}^{1/2} .
\end{equation} 

One then finds the following expressions for $C(d)$ corresponding to
four mutually-exclusive (and exhaustive) cases:
\begin{enumerate}
\item
If $d \ge 1 +a$ then the Sun and Moon's discs are separated and so 
\begin{equation}
\label{eqC1}
C(d) \equiv 0 .
\end{equation}
\item
If $1+a > d \ge \sqrt{\left| 1-a^2 \right|}$ then 
\begin{equation}
\label{eqC2}
C(d) = f(u) + a^2 f\left( {u\over a}\right) , 
\end{equation}
where 
\begin{equation}
\label{eqC3}
f(u) = {1 \over \pi} \; \Bigl( \arcsin u - u \sqrt{1 - u^2} \Bigr),
\end{equation}
and
\begin{equation}
\label{eqC4}
u = {1 \over 2 d} \left\{\left[ \left( 1 + a \right)^2 - d^2 \right] 
\left[ d^2 - \left( 1 - a\right)^2 \right] \right\}^{1/2} .
\end{equation}
Notice that these definitions ensure that $u$ is well-defined and
climbs monotonically from $u=0$ as $d$ decreases from $1+a$, not reaching
$u = \min(1,a)$ until $d = \sqrt{\left| 1-a^2 \right|}$. 
%
%
\item
If $\sqrt{\left| 1-a^2 \right|} > d \ge |1-a|$ then the expression for $C(d)$
depends on whether or not the Moon's apparent disc is larger than the Sun's.
\begin{enumerate}
\item
If $a \le 1$ then 
\begin{equation}
\label{eqC5}
C(d) = f(u) + a^2 \left[ 1 - f\left({u\over a}\right) \right].
\end{equation}
\item
If $a > 1$ then 
\begin{equation}
\label{eqC6}
C(d) = 1 - f(u) + a^2 f\left({u\over a}\right) .
\end{equation}
\end{enumerate}
\item
Finally, if $|1 - a| > d \ge 0$ then 
\begin{equation}
\label{eqC7}
C(d) = 1 \quad \hbox{if} \quad a \ge 1 
\qquad\qquad \hbox{and} \qquad\qquad 
C(d) = a^2 \quad \hbox{if} \quad a < 1 .
\end{equation}
\end{enumerate}

For neutrino physics we
require the distance $d_M$ traveled by the neutrino inside the Moon.
Defining the fraction $x =\displaystyle\frac {d_M} {2 R_M}$ where $R_M$ is the lunar radius, $x$
can  be given in terms of $C$ by the following formulae:
\begin{eqnarray}
x  &=& \sqrt {(4 z (2 - z) - 3)}\label{eq1} \\
C &=& \frac {2} {\pi} \left(\cos^{-1} (1- z) - (1- z) \sqrt{z (2 - z
)}\right) 
\label{eq2}
\end{eqnarray}
Eq.(\ref{eq2}) can be inverted to get the parameter $z$ for a given
$C$ and this $z$ can be substituted in eq.(\ref{eq1}). The
relationship between $x$ and $C$ so obtained is plotted in Fig\ref{Fig. 1}. When
the lunar disc passes through the center of the Sun, $C$ is 0.39 and
the neutrino eclipse starts at this value of $C$.When the optical
coverage increases above 39\%, $x$ rises sharply from zero and reaches
0.6 and 0.95 for optical coverage of 50\% and 80\% respectively

        For any point of observation of the usual solar eclipse (which
we shall call {\em single eclipse})  there
is a corresponding point on the other side of Earth where a double
eclipse occurs. With the coordinates labeled as (latitude, longitude), 
the single eclipse point ($\alpha,\beta$) is related to the double
eclipse point ($\lambda,\sigma$) by the relations (see Fig.\ref{Fig. 2}):
\begin{eqnarray}
\lambda&=&\alpha+2 \delta\nonumber \\
\sigma&=&\pi-2\:\Theta_{UT}-\beta 
\end{eqnarray}
where,
\begin{equation}
\sin \delta = \sin 23.5^o \sin (\frac{2\pi t}{T_{Y}}) ,
\end{equation}
$T_Y$ is the length of the year , zero of time t is chosen at midnight of
autumnal equinox {\it i.e.}Sept. 21, and $\Theta_{UT}$ is the angle
corresponding to the Universal Time -- UT.

During a double eclipse, the neutrinos  travel through Earth 
in addition. The distance $d_E$ traveled by the 
neutrino inside Earth along
the chord between the points $(\alpha,\beta)$ and
$(\lambda,\sigma$)  as a function of time $t$, is given by 
\begin{equation}
 d_E = 2 R_E (\sin \lambda \sin \delta + \cos \lambda \cos\delta \cos
(\frac{2\pi t}{T_D}))
\end{equation} 
where $R_E$ is the radius of Earth and $T_D$ is the length of the
day. This is the same distance that needs to be calculated in the
study of the day--night effect and a plot of this distance as a
function of $t$ is given in our earlier paper\cite{ours}.

 Present and upcoming high statistics neutrino detectors expect to
collect utmost a few solar neutrino events every hour.  As shown in
Sec.~\ref{calc}, single and double eclipse can lead to enhancements of
rates by upto two and a half times. Even with such large enhancements
during the eclipse the signal may not exceed statistical errors, since
each solar eclipse lasts only for a few hours. However they occur
fairly often. As many as 32 solar eclipses are listed to occur during
the 14 year period 1996 through 2010.  Global maps and charts are
available on the internet \cite{http}, or from computer programs
\cite{ephe}, for the location and duration of both the umbral and
penumbral coverage.
 In
future planning of neutrino--detector sites, these points may also be
kept in mind.
It is important to remark here that although
partial solar eclipses are not so useful to astronomers they can
nevertheless be relevant for neutrino physics, so long as $C$ is above
$0.39$.


We make the following comments concerning the neutrino possibilities
for the eclipses (single or double) that will be visible from
Kamiokanda, Sudbury or Gran Sasso (the site of Borexino) between 1997 and 2002:
%
\begin{description}
\item [March 9, 1997] -- This eclipse was a single eclipse at Kamiokanda, 
with an approximate duration of two and a half hours and a maximum 
optical coverage of just over 50\%. This eclipse was a double 
eclipse at the Gran--Sasso, for much the same duration, also with
roughly 50\%  optical coverage.
\item [February 26, 1998] -- For this eclipse only Gran Sasso saw
more than 39\% coverage, with the Moon and Sun's discs barely
touching as seen from Sudbury and 23\% coverage seen in Japan.
The eclipse was double in Gran--Sasso with a maximum coverage of 60\%. 
%
\item [August 11, 1999] -- {\em This eclipse will provide one of the best
opportunities}, being seen as a single eclipse with about 80\% 
optical coverage and almost 3 hour duration at Gran-Sasso. {\em It is also
visible from Sudbury}, with 70\% coverage, for just short of two hours 
close to the horizon near Sunrise. Unfortunately the Super--Kamiokanda site will 
get the smallest coverage, utmost about 30\%, for a double eclipse
lasting just over an hour.
%
\item [December 25, 2000] -- This eclipse lasts for three
hours around noon (and so is a single eclipse)
on Christmas day as seen from Sudbury, but the 
Moon's disc does not intersect the Sun's as seen from the other two sites. 
50\% is the maximum coverage presented to SNO.
\item [July 31, 2000] -- This eclipse is not visible at all from Japan, and
presents less than 10\% coverage in Italy, but is close to our threshold
of interest as seen from Sudbury, being a double eclipse seen from
there, with maximum coverage of 32\%. 
\item [December 14, 2001] -- Kamiokanda and Gran Sasso see respectably
large coverage this day (70\% and 85\%, respectively), with both seeing
a double eclipse whose duration is about 100 minutes. By contrast,
Sudbury sees only a grazing eclipse with coverage not reaching 10\%.
\item [June 10, 2002] -- This two-hour-long (single) eclipse falls 
just below threshold --- 32\% maximum coverage --- seen from Kamiokanda.
Sudbury again sees less than 10\% coverage, while no eclipse occurs
at all from Gran Sasso's perspective. 
\end{description}

Tables \ref{Table 1} through \ref{Table 3} list the times, durations,
maximum coverage and direction of all these eclipses.  Plots of the
optical coverage $C$ and the fractional distance $x$ traveled by the
neutrino as a function of time $t$ for the same eclipses are
given in Fig.\ref{Fig. A} to \ref{Fig. C}

The lesson to be learned from these tables and figures is that there
is, on the average, about one eclipse per year for which the maximum coverage
exceeds 39\%, and once several detectors become available
it will not be uncommon for any of these eclipses to be
seen by more than one detector at a time.

\section{Theory}
\label{th} 
\subsection{Regeneration in the Moon}
\label{Moon}
We now describe a straightforward way of obtaining the neutrino
regeneration effect in the Moon. The Moon has a radius of $1738$ km
and has an approximately uniform density of $3.33$gms/cc, except for a
core of radius $238$ km, where the density jumps to $7.55$
gm/cc\cite{Moonshadow}. The effect of the core can be treated by the
method in Sec.\ref{secMoonEarth}, but this is required only for
$x>0.99$ which occurs only very rarely, as seen from
Figs.(\ref{Fig. A}--\ref{Fig. C}). Hence we restrict ourselves to a model of Moon
of constant density.

Let a neutrino of flavor $\alpha$ be produced at time $t=t_0$ in the
core of the Sun.
Its  state vector is 
\begin{equation}
| \Psi_{\alpha} (t_0) \rangle = |\nu_\alpha \rangle = \sum_i U^S_{\alpha i}
| \nu_i^S \rangle.
\end{equation}
where $|\nu_i^S \rangle$ are the matter dependent mass eigenstates
with mass eigenvalues $\mu_i^S$ and
$U^S_{\alpha i}$ are the matrix elements of the matter dependent mixing
matrix in the core of the Sun. We use Greek index $\alpha$ to denote the three
flavors e, $\mu$,$\tau$ and Latin index i to denote the mass eigenstates
i = 1,2,3. The neutrino propagates in the Sun adiabatically upto $t_{R}$ (the
resonance point), makes non-adiabatic transitions at $t_R$, propagates
adiabatically upto $t_1$ (the edge of the Sun) and propagates as a free particle
upto $t_2$ when it enters the Moon. So the state vector at $t_2$ is
\begin{equation}
| \Psi_{\alpha} (t_2) \rangle 
= 
\sum_{j,i}  |\nu_j \rangle exp \left(-i \varepsilon
_{j} (t_{2}- t_{1}) \right)  exp \left(-i\int_{t_{R}}^{t_1} \varepsilon^{S}_{j} (t) dt 
\right) M^S_{ji} exp \left(-i \int_{t_0}^{t_R} \varepsilon^S_{i}(t) dt \right) 
U^S_{\alpha i}.
\label{psi2}
\end{equation}
where $ \varepsilon^S_{i} (t)(\equiv E+(\mu_i^S(t))^2/{2 E})$ are the matter dependent energy eigenvalues in the Sun,
$ \varepsilon_{i}$ and $|\nu_{i} \rangle$ are the energy eigenvalues and the 
corresponding eigenstates in vacuum and $M^S_{j i}$ is the probability amplitude for
the non-adiabatic transition $i\rightarrow j$. We multiply the right hand side of  
eq.(\ref{psi2}) by $ \sum_{k} |\nu^M_{k} \rangle \langle \nu^M_{k}|$ ( = 1)where 
$| \nu^M_{k}\rangle$ (i =1,2,3) is the complete set of matter dependent mass 
eigenstates inside the Moon. The neutrino propagates upto the the other end of the
Moon at $t_3$, and the  state vector at $t_3$ is
\begin{eqnarray}
| \Psi_{\alpha} (t_3) \rangle & = & \sum_{k,j,i} |\nu_{k}^M \rangle
exp 
\left( -i \varepsilon_{k}^M (t_{3}-t_{2})\right) 
\langle \nu_{k}^M|\nu_{j} \rangle 
exp 
\left( -i \varepsilon_{j} (t_{2}-t_{1}) -i \int_{t_R}^{t_1}
\varepsilon^S_{j} (t) dt \right)\nonumber\\  \nonumber
& &\times M^S_{j i} exp \left(-i \int_{t_0}^{t_R} \varepsilon^S_{i}
(t) dt \right) U^S_{\alpha i}\\  \nonumber 
&=& \sum_{k,j,i} |\nu_{k}^M \rangle
exp 
\left( -i \varepsilon_{k}^M (t_{3}-t_{2})\right) 
M^M_{k j} 
exp 
\left( -i \varepsilon_{j} (t_{2}-t_{1}) -i \int_{t_R}^{t_1}
\varepsilon^S_{j} (t) dt \right)\\  
& &\times M^S_{j i} exp \left(-i \int_{t_0}^{t_R} \varepsilon^S_{i}
(t) dt \right) U^S_{\alpha i}
\label{psi3}
\end{eqnarray}
We have introduced the probability amplitude $M^M_{k j}$ for non-adiabatic transitions
$j\rightarrow k$ due to the abrupt change in density when the neutrino enters the 
Moon . It is given by
\begin{equation}
  M^M_{k j} = \langle \nu_{k}^M| \nu_{j} \rangle = \sum_{\gamma} \langle \nu_{k}^M|
\nu_{\gamma} \rangle \langle \nu_{\gamma}| \nu_{j}\rangle
    =   \sum_{\gamma} U^M_{\gamma k} U^*_{\gamma j}
\label{noref}
\end{equation}
where $U_{\gamma j}$ is the mixing matrix in vacuum .
We multiply the right hand side of eq.(\ref{psi3}) by  $ \sum_{l} |\nu_ {l} \rangle \langle \nu_{l}|$ 
where  $| \nu_ {l}\rangle$ (l =1,2,3) is the complete set of vacuum mass eigenstates.
The neutrino leaves the other end of the Moon at $t = t_3$ and propagates upto the
surface of Earth ,which it reaches at $t_4$ .So the state vector at $t_4$ is
\begin{eqnarray}
| \Psi_{\alpha} (t_4) \rangle & = & \sum_{k,j,i,l} |\nu_{l} \rangle 
exp 
\left( -i \varepsilon_{k}^M (t_{3}-t_{2})\right) 
\langle \nu_{l}|\nu_{k}^{M} \rangle M^{M}_{k j} 
exp 
\left( -i \varepsilon_{j} (t_{2}-t_{1}) -i \int_{t_R}^{t_1}
\varepsilon^S_{j} (t) dt \right) \nonumber\\
& &\times M^S_{j i} exp \left(-i \int_{t_0}^{t_R} \varepsilon^S_{i}
(t) dt \right) U^S_{\alpha i} exp \left(-i \varepsilon_{l}
(t_{4}-t_{3})\right) \nonumber\\
&=& \sum_{k,j,i,l} |\nu_{l} \rangle
M^{M}_{k j} M^{M *}_{k l} M^{S}_{j i}
U^{S}_{\alpha i} exp \left (-i \Phi_{i j k l} \right)
\label{psi4}
\end{eqnarray}
where
\begin{equation}
\Phi_{i j k l} = 
\varepsilon_{k}^M (t_{3}-t_{2}) +\varepsilon_{l}(t_{4}-t_{3}) +\varepsilon_{j}
(t_{2}-t_{1}) + \int_{t_R}^{t_1} \varepsilon^{S}_{j} (t) dt + \int_{t_0}^{t_R}
\varepsilon^S_{i} (t) dt  
\end{equation}
We have used the fact that the the probability amplitude  for non-adiabatic 
transitions $k \rightarrow l$
 due to the abrupt change in density when the neutrino leaves the 
Moon   is 
\begin{equation}
 \langle \nu_{l}| \nu_{k}^{M} \rangle = M^{M*}_{k l}
\end{equation}
The probability of detecting a neutrino of flavor $\beta$ at $t_{4}$ is
\begin{equation}
|\langle \nu_\beta |\Psi_{\alpha} (t_4) \rangle|^2 = \\ \nonumber 
\sum \\ \nonumber
U^{*}_{\beta l} U_{\beta l'} M^M_{k j} M^{M*}_{k' j'} M^{M*}_{k l} M^{M}_{k' l'} M^S_{j i} M^{S*}_{
j' i'} U^S_{\alpha i} U^{S*}_{\alpha i'}  \\ \nonumber 
exp \left(-i(\Phi_{i j k l} -
\Phi_{i' j' k' l'} \right)
\end{equation}
where the summation is over  the set of indices $i,j,k,l,i',j',k',l'$ 
Averaging over $t_{R}$ leads to $\delta_{i i'} \delta_{j j'}$ and this
results in the desired incoherent mixture of mass eigenstates of
neutrinos reaching the surface of the Moon at $t_2$. Calling this
averaged probability as $P_{\alpha \beta}^M$ ( the probability for a
neutrino produced in the Sun as $\nu_\alpha$ to be detected as
$\nu_\beta$ in Earth after passing through the Moon), we can write the result as
\begin{equation}
 P_{\alpha \beta}^M = \sum_{j} P^S(\alpha \rightarrow j) P^M(j
\rightarrow \beta) 
\label{PabM}
\end{equation}
where
\begin{eqnarray}
 P^{S}(\alpha \rightarrow j) & = & \sum_i |M^S_{j i}|^2 |U^S_{\alpha
i}|^2  \\
P^M(j \rightarrow \beta) & = & 
\sum_{l,k,l',k'} U^{*}_{\beta l} U_{\beta l'} M^M_{k j} 
M^{M*}_{k' j} M^{M*}_{k l} M^{M}_{k' l'} exp \left(-i(\varepsilon^M_{k} - \varepsilon^M_{k'})
d_M - i(\varepsilon_{l} - \varepsilon_{l'}) r \right)
\end{eqnarray}
where we have replaced $(t_{3}- t_{2})$ by $d_{M}$, the distance traveled by the neutrino
inside the Moon, and $(t_{4} -t_{3})$ by $r$  the distance traveled by the neutrino 
from the Moon to Earth.
If there is no Moon , we put $d_M$ = 0, so that 
$P^M(j \rightarrow \beta)$ becomes $|U_{\beta j}|^2$ and so eq.(\ref{PabM})
reduces to the usual \cite{parke,narayan} averaged  probability 
for $\nu_{\alpha}$ produced in
the Sun to be detected as $\nu_{\beta}$ in Earth :
\begin{equation}
P^{O}_{\alpha \beta} = \sum_{i,j} |U_{\beta j}|^2 |M^S_{j i}|^2
|U^{S}_{\alpha i}|^2 .
\label{PabO}
\end{equation}

\subsection{Regeneration during double eclipse.}
\label{secMoonEarth}


We start
with $\Psi_\alpha(t_4)$ given by eq.(\ref{psi4}) and multiply the right hand side by $\sum_p |\nu_{p}^E \rangle \langle \nu_{p}^E |$
 (= 1) where $|\nu_{p}^E \rangle$ $(p = 1,2,3)$  is a complete set of
mass eigenstates just below the surface of Earth. If the neutrino
that enters Earth at time $t=t_{4}$ propagates adiabatically upto
$t_5$ (non-adiabatic jumps during the propagation will be considered
subsequently), its state vector at time $t=t_5$ is 
\begin{equation}
| \Psi_{\alpha} (t_5) \rangle  =  \sum_{k,j,i,l,p} |\nu_{p}^E \rangle
M^{E}_{p l} M^{M}_{k j} M^{M *}_{k l} M^{S}_{j i}
U^{S}_{\alpha i}  \exp \left (-i \Phi_{i j k l p } \right)
\end{equation}
where
we have introduced the probability amplitude $M^E_{p l}$ for non adiabatic transitions
$l\rightarrow p$ due to the abrupt change in density when the neutrino enters the 
Earth . It is given by
\begin{equation}
 M^E_{p l} = \langle \nu_{p}^E| \nu_{l} \rangle = \sum_{\sigma}
 U^E_{\sigma p} U^*_{\sigma l}
\label{mekj}
\end{equation}
and
\begin{equation}
\Phi_{i j k l p } = \int_{t_4}^{t_5}
\varepsilon_{p}^{E} dt+\varepsilon_{l}(t_{4}-t_{3}) +
\varepsilon_{k}^M (t_{3}-t_{2}) + \varepsilon_{j} (t_{2}-t_{1}) +
\int_{t_R}^{t_1} \varepsilon^{S}_{j} (t) dt + \int_{t_0}^{t_R} 
\varepsilon^S_{i} (t) dt  
\end{equation}
The probability of detecting a neutrino of flavor $\beta$ at $t_{5}$ is
\begin{eqnarray}
|\langle \nu_\beta |\Psi_{\alpha} (t_5) \rangle|^2 &=& 
\sum 
U^{E*}_{\beta p} U^{E}_{\beta p'} M^{E}_{p l} M^{E*}_{p' l'} 
M^M_{k j} M^{M*}_{k' j'} M^{M*}_{k l} M^{M}_{k' l'} M^S_{j i} M^{S*}_{
j' i'} U^S_{\alpha i} U^{S*}_{\alpha i'} \nonumber \\ 
&&\times\exp \left(-i(\Phi_{i j k l p} -
\Phi_{i' j' k' l' p'} \right)
\end{eqnarray}
where the summation is over  the set of indices $i,j,k,l,p,i',j',k',l'p'$ 
Again averaging over $t_{R}$  and
calling this
averaged probability as $P_{\alpha \beta}^{M E}$ ( the probability for a
neutrino produced in the Sun as $\nu_\alpha$ to be detected as
$\nu_\beta$ in Earth after passing through the Moon and Earth), we can write the result as
\begin{equation}
P_{\alpha \beta}^{M E} = \sum_{j} P^S(\alpha \rightarrow j) P^{M E} (j
\rightarrow \beta) 
\label{PabME}
\end{equation}
where
\begin{eqnarray}
P^{M E}(j \rightarrow \beta) & = & 
\sum_{l,k,p,l',k',p'} U^{E*}_{\beta p} U^{E}_{\beta p'} 
M^E_{p l} M^{E*}_{p' l'}  
M^M_{k j} M^{M*}_{k' j} M^{M*}_{k l} M^{M}_{k' l'} \nonumber \\
&&\times exp \left(-i\int_{t_4}^{t_5}
(\varepsilon_p^E(t)-\varepsilon_{p\prime}^E(t)) dt
-i(\varepsilon^M_{k}-\varepsilon^M_{k'}) d_{M}
- i(\varepsilon_{l} - \varepsilon_{l'}) r \right)
\label{MoonEarth}
\end{eqnarray}
where we have replaced 
 $(t_{3}- t_{2})$ by $d_{M}$, the distance traveled by the neutrino
inside the Moon and $(t_{4} -t_{3})$ by $r$  the distance traveled by the neutrino 
from the  Moon to  Earth.

We next show how to take into account non-adiabatic jumps during the
propagation inside Earth. Consider $\nu$ propagation through a
series of slabs of matter, density varying inside each slab smoothly
but changing abruptly at the junction between adjacent slabs. The
state vector of the neutrino at the end of the $n^{th}$ slab $|n
\rangle$ is related to that at the end of the $(n-1)^{th}$ slab $|n-1
\rangle$ by $|n \rangle = F^{(n)} M^{(n)} |n-1 \rangle$ where $M
^{(n)}$ describes the non-adiabatic jump occuring at the junction
between the $(n-1)^{th}$ and $n^{th}$ slabs while $F^{(n)}$ describes
the adiabatic propagation in the $n^{th}$ slab. They are given by
\begin{eqnarray}
M^{(n)}_{i j} &=& \langle \nu_{i}^{(n)} | \nu_{j}^{(n-1)} \rangle  = (U^{(n)^ {\dagger}}
U^{(n-1)})_{i j}^*  \, , \label{Mn}\\
F^{(n)}_{i j} &=& \delta_{i j} exp \left(-i \int_{t_{n-1}}^{t_{n}} \varepsilon_{i}
(t) dt \right)\, ,
\end{eqnarray}
where the indices $(n)$ and $(n-1)$ occuring on $\nu$ and $U$  refer 
respectively to the $n^{th}$ and $(n-1)^{th}$ slabs at the junction between these 
slabs. Also note that $M^{(1)}$ is the same as $M^{E}$ defined in eq.(\ref{mekj}). Defining
the density matrix at the end of the $n^{th}$ slab as
$\rho^{(n)} = | n \rangle \langle n |$ , we have the recursion formula
\begin{equation}
\rho^{(n)} = F^{(n)} M^{(n)} \rho^{(n-1)} M^{(n)\dagger}
F^{(n)\dagger}.
\label{rho}
\end{equation} 

Corresponding to the state vector $|\Psi_\alpha(t_4)\rangle$
(eq.\ref{psi4}) of the $\nu$ entering Earth, the density matrix is
$|\Psi_\alpha(t_4)\rangle \langle\Psi_\alpha(t_4)|$. After averaging
over $t_R$, we call it $\rho^{(0)}$ and 
calculate $\rho^{(N)}$ recursively at the end of the $N^{th}$ slab using (\ref{rho}).
The probability of observing $\nu_\beta$ at the end of the $N^{th}$ slab is
\begin{equation}
P^{ME}_{\alpha \beta} = \langle \nu_{\beta} | \rho^{(N)} | \nu_{\beta} \rangle = ( U^{(N)}
\rho ^{(N)^*}  U^{(N)^ {\dagger}} )_{\beta \beta} \, .
\end{equation}
This formula (which reduces to eq.(\ref{PabME}) and (\ref{MoonEarth})
for $N$ = 1) can be 
used for Earth modeled as consisting of $(N+1)/2$ concentric
shells, with the density varying gradually within each shell. For the
Earth, the major discontinuity in the density occurs at the boundary
between mantle and core and adequate accuracy can be achieved with
$N=3$ (mantle and core). However, for $(d_E/2 R_E)<0.84$ neutrinos
pass only through mantle and so, $N=1$.

For the sake of completeness, we state that if we put $d_{M}$ = 0 in
eq.(\ref{MoonEarth}) 
we get the regeneration in Earth alone:
\begin{equation}
P_{\alpha \beta}^E = \sum_{j} P^S(\alpha \rightarrow j) P^E(j
\rightarrow \beta) 
\label{PabE}
\end{equation}
where
\begin{equation}
P^E(j \rightarrow \beta)  =  
\sum_{k,k'} U^{E*}_{\beta k} U^{E}_{\beta k'} M^E_{k j} 
M^{E*}_{k' j} exp \left(-i\int_{t_4}^{t_5}
(\varepsilon_p^E(t)-\varepsilon_{p\prime}^E(t)) dt
\right)
\label{Pejb}
\end{equation}
Eqs.(\ref{PabE}) and (\ref{Pejb}) have been used to study the day--night effect \cite{ours}.

It is important to note that the factorization of probabilities seen
in eqs(\ref{PabM}),(\ref{PabME}) and (\ref{PabE}) is valid only for
mass eigenstates in the intermediate state. An equivalent statement of
this result is that the density matrix is diagonal only in the
mass-eigenstate representation and not in the flavour representation.

\subsection{Three flavor mixing parameters}
We parameterize the mixing matrix U in vacuum as
$U = U^{23} (\psi) U^{13} (\phi) U^{12} (\omega)$
where $U^{ij} (\theta_{ij})$ is the two flavor mixing matrix between the $i^{th}$ and the
$j^{th}$ mass eigenstates with the mixing angle $\theta_{ij}$, neglecting CP violation.
In the solar neutrino problem $\psi$ drops out \cite{ajmm,kuopan}  
The mass differences in
vacuum are defined as $\delta_{21} = \mu^{2}_{2} - \mu_{1}^{2}$ and
$\delta_{31} = \mu^2_{3} - \mu^2_{1}$. It has been shown \cite{narayan,fogli} 
that the
simultaneous solution of both the solar and the atmospheric neutrino
problems requires
\begin{equation}
\delta_{31} \gg \delta_{21}
\label{delgt}
\end{equation}
 and under this condition $\delta_{31}$ also drops out.
The rediagonalization of the mass matrix in the presence of matter (in
solar core, Moon or
Earth) under condition (\ref{delgt}) leads to the following results \cite{narayan}
\begin{eqnarray}
\tan 2 \omega_m  &=&  \frac{\delta_{21} \sin 2 \omega}{
                  \delta_{21} \cos 2 \omega - A \cos^2 \phi} \label{eq13} \\ 
\sin \phi_{m} &=& \sin \phi  \label{eq14} \\
\delta_{21}^{m} &=& \delta_{21} \cos 2 (\omega - \omega_{m})- A \cos^2\phi \cos 2 
\omega_{m} 
\label{eq15}
\end{eqnarray} 
where A is the Wolfenstein term $A = 2 \sqrt{2}~G_F~N_e\,E$ ($N_{e}$
is the number density of electrons and E is the neutrino energy)
. We note that $\delta_{31}\gg A^S, A^M, A^E$. In
eqs.(\ref{eq13})--(\ref{eq15}) the subscript ``$m$'' stands for
matter. Under the condition $\delta_{31}\gg A\approx\delta_{21}$ we
need the non adiabatic transition probability $|M^{S}_{ij}|^2$ for i,j
=1,2 only and this is taken to be the modified Landau--Zener jump
probability for an exponentially varying solar density
\cite{kuopan}.

\section{Calculations and Results}
\label{calc}
The neutrino detection rates for a Super--Kamoika type of  detector is
given by
\begin{equation}
R =\int\phi(E)\,\sigma (E) P_{ee} dE +
\frac{1}{6} (\int\left( \phi (E) \sigma (E) (1-P_{ee}) dE \right) 
\end{equation}
where the second term 
is the  neutral current contribution for $\nu_\alpha(\alpha\ne e)$ and $\phi(E)$ is the solar
neutrino flux as a function of the neutrino energy $E$ and
$\sigma(E)$ is the cross section from neutrino electron scattering and
we integrate from  $5MeV$ onwards. The
cross section is taken from  \cite{bahcall} and the flux from
\cite{bahpin92}. The rates for a single eclipse, double eclipse and
without eclipse (at day--time) $R_{M},R_{ME}$ and $R_{O}$  are calculated
using $P_{ee}^M,P_{ee}^{ME}$ and $P_{ee}^O$ from
eqns.(\ref{PabM}),(\ref{PabME}) and (\ref{PabO}) respectively. 
We define the enhancement factors $F$ and $G$ for a single and double
eclipse respectively:
\begin{eqnarray}
F &=& \frac{R_{M} - R_{O}} {R_{O}}\\
G &=& \frac{R_{ME} - R_{O}} {R_{O}}.
\end{eqnarray}

It is easy to see that $F$ and $G$ have to be less than 5 and this
theoretical maximum value occurs when $P_{ee}^O=0$ and $P_{ee}^M$ and
$P_{ee}^{ME}$ are put 1. If one imposes the constraint that the
observed \cite{mohanetal} neutrino rate is $0.51\pm 0.07$ of the
prediction of the 
standard solar model, the maximum possible  enhancement is
reduced to about 1.40 (at 90\% C.L.).

Although we have given the theory for double eclipse in
Sec.\ref{secMoonEarth}, we shall restrict the detailed calculations to
the single eclipse in the present paper. For the double eclipse we
present results only for the model of Earth with  constant density
($5.52$ gms/cc). This can be taken as a rough guide and detailed
calculations for the double eclipse are reserved for the future. 

We calculate the enhancement
factors $F$ and $G$  for various values of the neutrino
parameters, $\omega$, $\delta_{21}$, and $\phi$.
We show the results as  contour plots in the $\delta_{21}$--$\omega$ plane for
different values of  $\phi$. Figs.\ref{Fig. 3} and \ref{Fig. 4} show
the  $F$--contours for $\phi=0^o$ and $\phi=30^o$ respectively. 
For each $\phi$ we show the contours for different
distances of travel of the neutrino through the Moon.
Fig. \ref{Fig. 5} shows the $G$--contours for $\phi=0$ for the maximum
distance of travel of the neutrino inside the Moon and Earth.
The main features of the results are as follows:

\begin{itemize}
\item As the distance traveled by the neutrino inside the Moon increases one can
see an appreciable increase in the enhancement factor $F$. It increases from less than 
$10\%$ to about almost $100\%$
when the neutrino travels the whole diameter of
the Moon in the case of two flavor mixing i.e $\phi$ = $0$.
\item Large  ($> 40\%$) values of $F$ occur for
$\omega$ between $20^o$ and $30^o$ and $\delta_{21} \sim 10^{-6} eV^2$
This is true even for nonzero $\phi$.
\item It may be noted from Fig.\ref{Fig. 3} that the maximum enhancement
region for $\phi=0$ is not far away from the present (2--generation) best fit
parameters of Super Kamiokanda \cite{ICHEP}($\delta_{21}=1.4\times10^{-7} eV^2;
\omega =22^o$), which may be encouraging for the observation of
the eclipse effect.
\item The effect of a non zero ``13'' mixing angle $\phi$, is to
dilute the enhancement 
factor $F$ for all values of distance traveled through the Moon.(In
fact for $\phi \approx 
45^o$, $F$ is practically zero and so we have not plotted this case.)
This is because
a non zero $\phi$ means $\nu_{e} \leftrightarrow \nu_{\tau}$ oscillations,
and matter cannot  reconvert $\nu_{\tau}$ back to $\nu_{e}$, because the
``13'' mixing angle $\phi$ is not affected by matter.
\item If large enhancement $F$ is seen for values of $x \leq 0.6$, it immediately
signals a very small value of $\phi$. On the other hand, if no
enhancement is seen for 
small $x$ but  there is enhancement
only for $x \geq 0.8$ it signals an appreciable value of $\phi$.
\item For a double eclipse there are considerable enhancements even for small
values of $\omega$.  There is enhancement throughout the range of $\omega$
from small angles till about $40^o$. In fact the regions of largest
enhancement ($>100\%$)  
are for $\omega$ between $5^o$ to $20^o$. Although  the region of
maximum enhancement factor $G$ is centered around $\delta_{21}\sim
3\times 10^{-6}eV^2$,
sizeable enhancement occurs over a wide range of
$\delta_{21}$. However, the details are likely to change when
calculations are done for a more realistic density distribution inside
Earth.
\item If enhancement is not seen, then certain regions of the neutrino
parameter space can be excluded.If no enhancement is seen for single
eclipse, a panel of $\omega$ between $5-25^o$ and $\delta_{21} \approx
2 \times 10^{-7}-2 \times 10^{-6} eV^2$ for $\phi$ = 0 can be ruled
out. If it is not seen for a double eclipse, a larger region can be
ruled out.
\end{itemize}

Having calculated the enhancements as a function of the distance
traveled, we may now estimate roughly the enhancement in the event
rate for an eclipse at a particular site. For this purpose we choose
an eclipse with a rather long duration to evaluate the
enhancement. One such eclipse occurs rather soon at Gran Sasso on
August 11, 1999 with 84 minutes of duration with nonzero $x$. A simple
estimate based on the enhancements presented in Fig.\ref{Fig. 3}
integrated using the time--dependence presented in Fig.\ref{Fig. C}
for this eclipse, leads one to conclude that the total neutrino count
during this 84 minutes could be enhanced by a factor of 1.5 for
$\phi=0$, $\omega=22^o $, $\delta_{21}=10^{-6}eV^2$ and it is less for
other values of the neutrino parameters.

Our calculations thus indicate that the observation of the eclipse
effect is {\em not} possible at the present detectors as the counting
rates currently are no more than about one per hour. However, the
calculated enhancements are not too small to be discarded
completely. We envisage that the eclipse effect can become detectable
in the near future in two or three ways. Experimental neutrino physics
will continue to cross new frontiers with innovative techniques
leading to counting rates that are larger by an order of
magnitude. One such proposal that has been already made is the Borex
detector\cite{Borex} where the counting rate can be as large as 40 per
hour.  With such a detector, our calculations of enhancement factors
show that the eclipse effect would be clearly observable. Another
possibility is the accumulation of data during a large number of
eclipses to obtain a statistically significant sample, which may enable
one to observe an
enhancement or to rule out a given parameter space. Yet another
interesting possibility is to exploit the fact that some eclipses
occur simultaneously at two sites. Correlation between the data
collected at the two sites can enhance the statistical significance of
scanty data. 

It must be remembered that the counting rates at the present--day
detectors such as Super-Kamiokanda could hardly have been anticipated 20 years
back.  One of the strengths of neutrino physics is that even when
counting rates have been small the accumulated effect has stood the
test of time over a 30 year period.  The eclipse effect is perhaps an
effect which may be observable only in a prolonged study and ours is
only an effort to initiate such a study. It is also pertinent to remark
here that although the original proposal to observe the shadow of the
Moon\cite{referee} in high-energy cosmic rays was made in
$1957$\cite{shadows}, it took more than 30 years to observe it\cite{shadow2}.


\section{Discussion}
\label{disc}
We have studied the effect on the solar neutrinos of their passage through
the Moon as well as the Moon together with Earth. Although the
numerical results presented in the paper cover only a representative 
sample of the set of various parameters, our analytical expressions can be
used for more extensive calculations depending on the requirement. Also
one can go beyond the hierarchy : $\delta_{31} >> \delta_{21}$.

We now offer a few concluding remarks:
\begin{itemize}
\item Together with the day-night effect, the eclipse effects provide us
with the tools for studying solar neutrinos, in a way independent of
the uncertainties of the solar models.
\item If the neutrino mass differences are really very small
($\delta_{21} < 10^{-5} eV^2$) there is no way of pinning down the neutrino
parameters  other than using the astronomical objects such as the Moon or
Earth for the "long-base-line experiments".
\item It is important to stress that even the demonstrated absence of any
eclipse effect would provide us with definitive information on neutrino
physics.
\item Accumulation of data over many eclipses may be needed for good statistics.
In any future planning of detector sites, this may be kept in mind.
\item It appears that Nature has chosen the neutrino parameters in such a
way that the Sun affects the propagation of solar neutrinos. It may be hoped
that Nature has similarly chosen "lucky" parameters so that the Moon and the
Earth too can affect  the neutrinos!
\item Finally, we stress the novelty of the whole phenomenon, and urge the
experimentalists to look for and study the eclipse effects in an unbiased 
manner. They may even discover some surprises, not predicted by our
calculations!
\end{itemize}

Acknowledgments: We thank KVL Sarma for bringing ref \cite{Sher} to our attention,
N.D. Hari Dass for raising the possibility of detecting the neutrinos on the opposite
side of Earth during the eclipse, and M.C Sinha for much help with the astronomical
aspects of the problem and Sandip Pakvasa for useful discussions.

\begin{figure}
\caption{
 The fractional distance traveled by the neutrino inside the Moon
$x (= \displaystyle \frac {d_{M}} {2 R_{M}}$) is plotted against the optical
coverage $C$ of the solar eclipse.
}\label {Fig. 1}
\end{figure}
\begin{figure}
\caption{Geometry relating the double eclipse point $(\lambda,\sigma)$
to the single eclipse point $(\alpha,\beta)$. $(a)$ Section of Earth passing
through $(\alpha,\beta)$ and perpendicular to the ecliptic. $(b)$
Section passing through $(\alpha,\beta)$ and parallel to the equator.
}\label {Fig. 2}
\end{figure}
\begin{figure}
\caption{The fractional distance travelled by the neutrino inside the
Moon $(x)$ and the optical coverage $(C)$ are plotted as a function of
time t (in minutes), for five eclipses visible from Kamiokanda.}\label {Fig. A}
\end{figure}
\begin{figure}
\caption{
Same as Fig.~\protect\ref{Fig. A}, for six eclipses visible from
Sudbury.}
\label {Fig. B}
\end{figure}
\begin{figure}
\caption{Same as Fig.~\protect\ref{Fig. A}, for five eclipses visible from Gran Sasso.
}\label {Fig. C}
\end{figure}
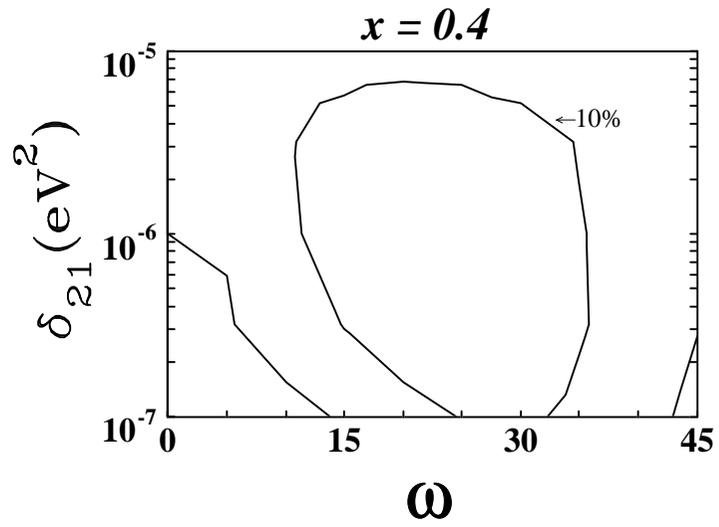
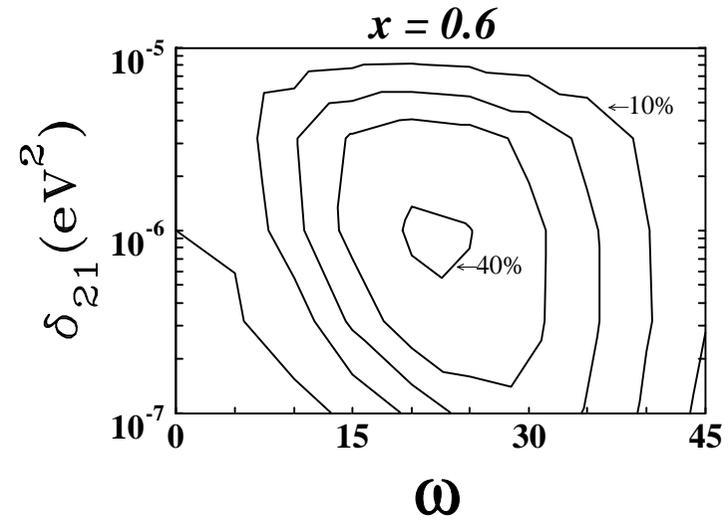
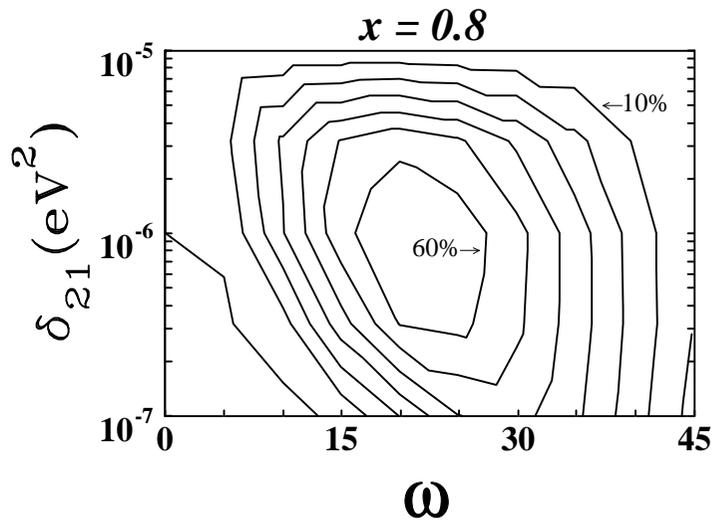
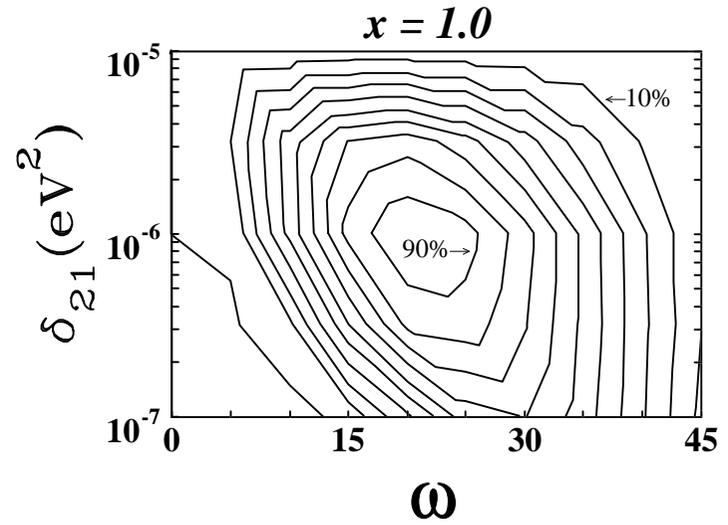
\begin{figure}
\caption{
 Contour plots of the enhancement factor for single eclipse
$F( = \displaystyle\frac {R_{M} - R_{O}} {R_{O}})$ in the $\omega - \delta_{21}$ plane
for $\phi = 0^o$ and for four values of $x$ ($x$ = 0.4, 0.6, 0.8 and 1.0).
The enhancement factor (regarded as a percentage) increases by $10\%$ for
every adjacent ring , as we move inwards towards the center of the plot.
}\label {Fig. 3}
\end{figure}
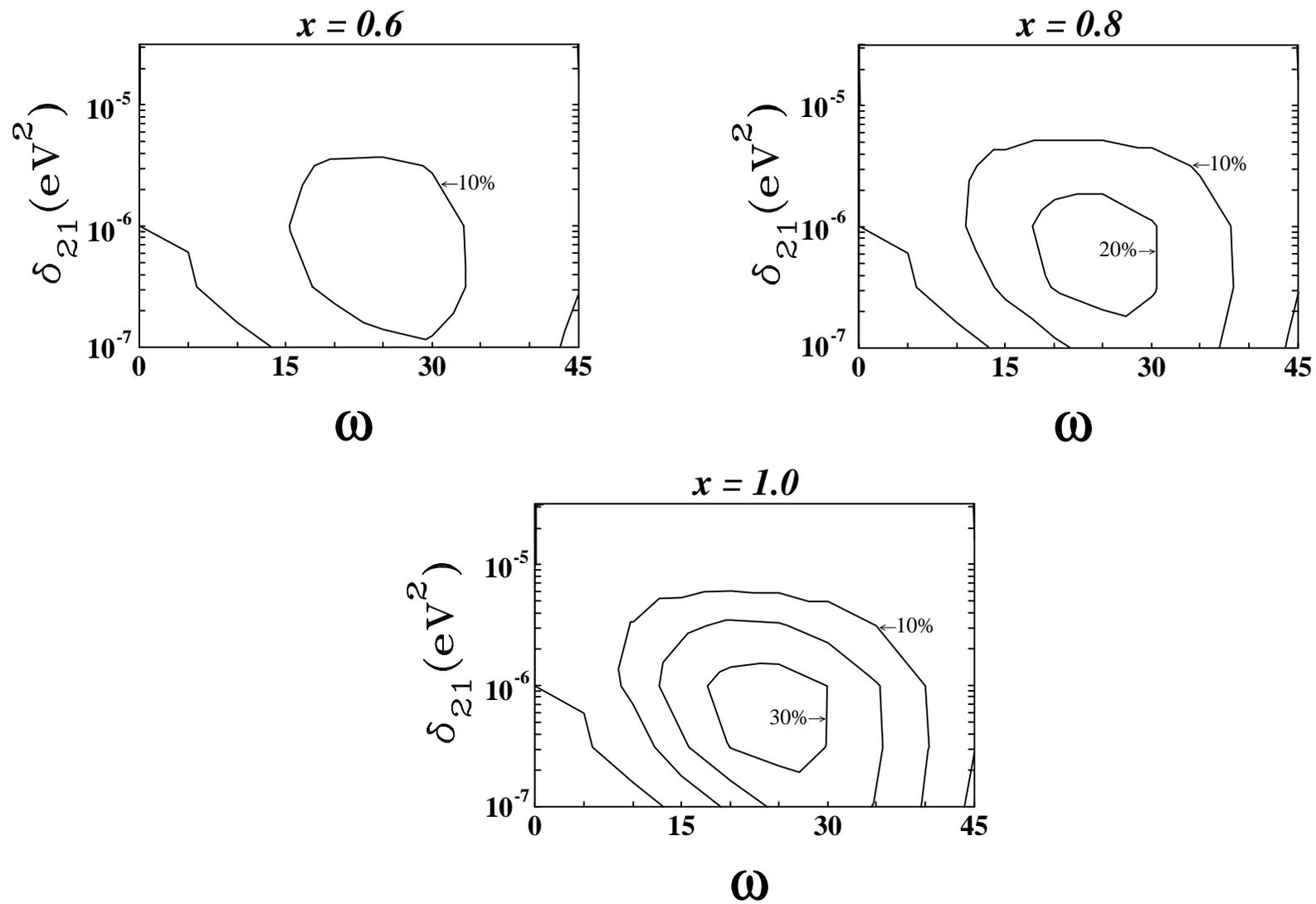
\begin{figure}
\caption{
 Contour plots of the enhancement factor for single eclipse
$F( = \displaystyle\frac {R_{M} - R_{O}} {R_{O}})$ in the $\omega - \delta_{21}$ plane
for $\phi = 30^o$ and  $x$ = 0.6, 0.8 and 1.0.
The enhancement factor (regarded as a percentage) increases by $10\%$ for
every adjacent ring , as we move inwards towards the center of the plot.
}\label {Fig. 4}
\end{figure}
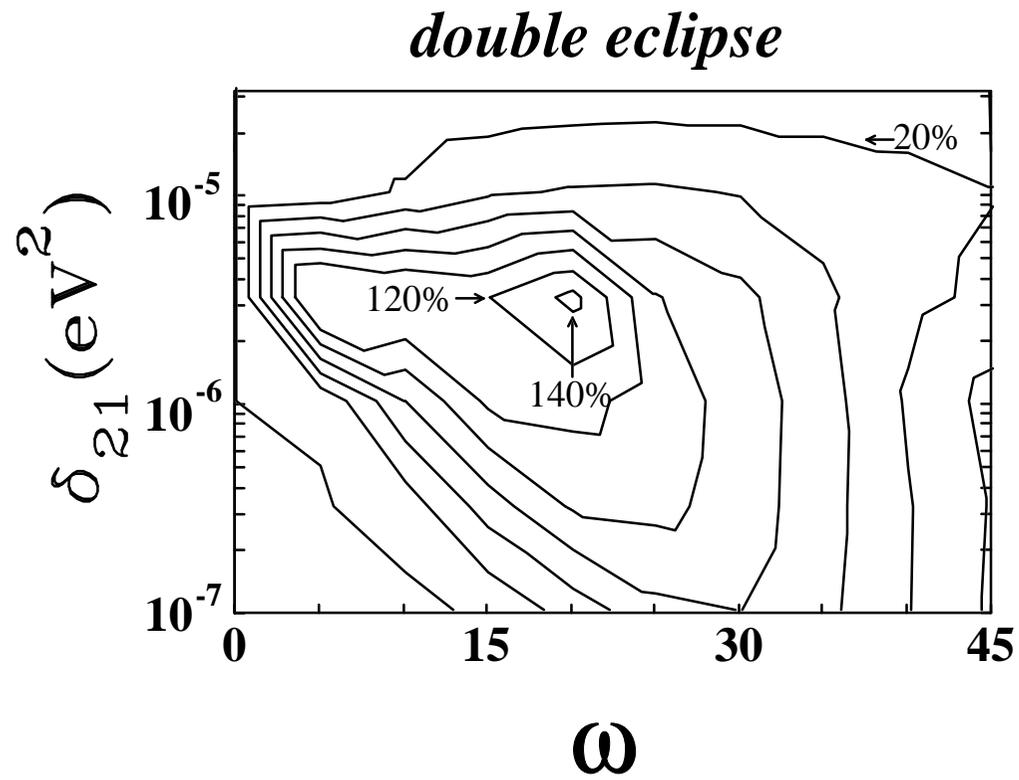
\begin{figure}
\caption{
Contour plot of the enhancement factor for double eclipse  $G (=  \displaystyle\frac {R_{M E}- R_{O}} {R_{O}}$)
for $\phi$ = 0 and $x$= 1.0. The distance travelled by the neutrino inside the
Earth is also taken to be the full Earth diameter. The enhancement factor
increases by $20\%$ as we move inwards. 
}\label {Fig. 5}
\end{figure}

\pagebreak
\def\deg{${}^o$}

\begin{table}
\begin{tabular}{cccc|cc|cccc|c}
Date && UT && Duration  ($T$) && \multicolumn{3}{c}{Solar Position}
&& $C$ (at max) \\ 
 &&   && && altitude && azimuth &&  \\ \hline
Mar. 9, 1997 && 1:06 && 143 min && 42\deg 16' &&
143\deg 28' && 0.53 \\ 
Feb. 26, 1998 && 16:57 && 67 min && -50\deg 25' 
&& 53\deg 07' && 0.23 \\ 
Aug. 11, 1999 && 11:46 && 74 min && -23\deg 11' 
&& 312\deg 17' && 0.30 \\ 
Dec. 14, 2001 && 19:16 && 108 min && -28\deg 56' 
&& 98\deg 47' && 0.71 \\ 
Jun. 10, 2002 && 22:37 && 122 min && 36\deg 33' 
&& 86\deg 37' && 0.32 \\ 
\end{tabular}
\vspace{1cm}
\caption{Eclipse parameters for eclipses visible from 
Super Kamiokanda (latitude: 36${}^o$ 24' N, longitude: 140${}^o$ 0' E) 
between 1997 and 2002. Date, Universal Time, the Sun's position
(given as altitude and azimuth from the observer's position, with negative
altitude indicating a double eclipse since the direction is below the horizon)
and Optical Coverage ($C$) are all given for the instant of maximum coverage. 
Eclipse Duration is given in minutes.}\label{Table 1}
\end{table}

\begin{table}
\begin{tabular}{cccc|cc|cccc|c}
Date && UT && Duration  ($T$) && \multicolumn{3}{c}{Solar Position}
&& $C$ (at max) \\ 
 &&   && && altitude && azimuth &&  \\ \hline
Mar. 9, 1997 && 2:36 && 58 min && -33\deg 04' &&
303\deg 19' && 0.13 \\ 
Feb. 26, 1998 && 18:08 && 31 min && 34\deg 28' 
&& 189\deg 02' && 0.003 \\ 
Aug. 11, 1999 && 9:40 && 100 min && -6\deg 28' 
&& 59\deg 34' && 0.72 \\ 
Dec. 25, 2000 && 17:29 && 189 min && 20\deg 07' 
&& 181\deg 07' && 0.51 \\ 
Jul. 31, 2000 && 2:45 && 72 min && -15\deg 19' 
&& 319\deg 24' && 0.32 \\ 
Dec. 14, 2001 && 21:57 && 89 min && -3\deg 43' 
&& 239\deg 35' && 0.09 \\ 
Jun. 11, 2002 && 1:05 && 59 min && 0\deg 43' 
&& 303\deg 45' && 0.08 \\ 
\end{tabular}
\vspace{1cm}
\caption{Eclipse parameters for eclipses visible from 
Sudbury (SNO) (latitude: 46${}^o$ 29' N, longitude: 81${}^o$ 0' W) 
between 1997 and 2002. Date, Universal Time, the Sun's position
(given as altitude and azimuth from the observer's position, with negative
altitude indicating a double eclipse since the direction is below the horizon)
and Optical Coverage ($C$) are all given for the instant of maximum coverage. 
Eclipse Duration is given in minutes.}\label{Table 2}
\end{table}

\begin{table}
\begin{tabular}{cccc|cc|cccc|c}
Date && UT && Duration  ($T$) && \multicolumn{3}{c}{Solar Position}
&& $C$ (at max) \\ 
 &&   && && altitude && azimuth &&  \\ \hline
Mar. 9, 1997 && 1:09 && 78 min && -44\deg 34' &&
41\deg 15' && 0.49 \\ 
Feb. 26, 1998 && 18:57 && 93 min && -23\deg 47' 
&& 280\deg 32' && 0.59 \\ 
Aug. 11, 1999 && 10:44 && 171 min && 62\deg 12' 
&& 165\deg 46' && 0.83 \\ 
Jul. 31, 2000 && 1:14 && 45 min && -23\deg 07' 
&& 31\deg 03' && 0.07 \\ 
Dec. 14, 2001 && 21:20 && 96 min && -61\deg 34' 
&& 304\deg 37' && 0.85 \\ 
\end{tabular}
\vspace{1cm}
\caption{Eclipse parameters for eclipses visible from 
Gran Sasso (latitude: 42${}^o$ 29' N, longitude: 13${}^o$ 30' E) 
between 1997 and 2002. Date, Universal Time, the Sun's position
(given as altitude and azimuth from the observer's position, with negative
altitude indicating a double eclipse since the direction is below the horizon)
and Optical Coverage ($C$) are all given for the instant of maximum coverage. 
Eclipse Duration is given in minutes.}\label{Table 3}
\end{table}

\end{document}